\documentclass[%
 reprint,
 amsmath,amssymb,
 aps,prc
]{revtex4-2}
\usepackage[colorlinks, citecolor=red]{hyperref}
\usepackage{graphicx}
\usepackage{dcolumn}
\usepackage{bm}
\usepackage{nameref}
\usepackage{natbib}
\usepackage[T1]{fontenc}
\usepackage{booktabs, array, mathptmx, float, tabularx, booktabs, lipsum, amsmath,multirow}
\usepackage{siunitx, xcolor}
\usepackage[version=4]{mhchem}

\begin{document}
\preprint{APS/123-QED}
\title{\texorpdfstring{$\alpha$}{}-decay for superheavy nucleus: The alpha decay energy to the one-fourth power }

\author{ Jinyu Hu$^{1}$ and  Chen Wu$^{1}$ } \affiliation{
\small 1. Xingzhi College, Zhejiang Normal University, Jinhua, 321004, Zhejiang, China}
\begin{abstract}
Recently, Sobhani and Luo \cite{sobhani2025unified} proposed a new empirical formula for $\alpha$ decay based on the $Q_\alpha^{-1/4}$ energy dependence, incorporating the proton number $Z$, neutron number $N$, and relative neutron excess $I = (N - Z)/(N + Z)$ as primary parameters. In this work, we extend this model by explicitly including the angular momentum of the emitted $\alpha$ particle and the quadrupole deformation of the daughter nucleus. Using this improved formula to evaluate the $\alpha$-decay half-lives of 400 nuclei yields a root-mean-square (RMS) deviation of 0.97 relative to experimental data. Furthermore, we employ support vector regression (SVR)-taking $Q_\alpha^{-1/4}$, $N$, $Z$, angular momentum, and daughter-nucleus deformation as input features-which further reduces the RMS deviation to 0.56. Finally, we apply both the extended formula and the SVR model to predict the $\alpha$-decay half-lives of even-even nuclei with $Z = 120$ and $Z = 122$. The predicted half-lives show good consistency with those from the Sobhani and Poenaru formulas, and both approaches strongly support $N = 184$ as the next neutron magic number.
\end{abstract}

\maketitle

\section{\label{sec:level1}INTRODUCTION}
Over the past two decades, significant progress in experimental facilities and techniques has allowed the laboratory synthesis of various superheavy nuclei(SHN), which predominantly undergo $\alpha$ decay. As an effective probe, $\alpha$ decay provides vital insights into the shell structure, ground-state energies, energy-level schemes, and nuclear spins of these extreme systems. What's more, The $\alpha$ deacy chains can provied new SHN elements information. As a result, $\alpha$-decay research continues to be a central domain in modern nuclear physics.\cite{qu2014comparative,zdeb2013half,sun2017systematic,santhosh2018alpha,hosseini2019alpha,oganessian2010synthesis,oganessian2007heaviest,hosseini2017theoretical,akrawy2019alpha,hosseini2019alpha,javadimanesh2013investigation,hosseini2019alpha,hodgson2003cluster}. 
\par{}
In 1908, Rutherford first observed $\alpha$ decay, in which a parent nucleus emits an $\alpha$ particle (a helium nucleus). In 1918, Geiger and Nuttall \cite{geiger1911lvii} established the empirical systematic law for $\alpha$-decay half-lives, demonstrating that the logarithm of the half-life scales linearly with the inverse square root of the decay energy. Subsequently, Gamow, as well as Condon and Gurney independently, described $\alpha$ decay as a quantum tunneling process wherein the $\alpha$ particle penetrates the potential barrier formed by its Coulomb interaction with the daughter nucleus. This picture naturally accounted for the Geiger–Nuttall law and provided compelling evidence for the validity of quantum mechanics.
\par{}
Subsequently, building on the foundational work of Geiger and Nuttall, Gamow\cite{gamow1928quantentheorie}, and Condon and Gurney\cite{gurney1928wave}, two primary approaches have been developed to calculate nuclear $\alpha$-decay half-lives. The first, rooted in the Geiger–Nuttall law, incorporates systemic properties across a broader range of nuclei and has yielded numerous empirical formulas. Representative examples include the Viola–Seaborg–Sobiczewski formula\cite{viola1966nuclear}, where the Geiger–Nuttall slope parameter is expressed as a linear function of the proton number; the Royer formula\cite{royer2000alpha}, derived from quantum tunneling principles; and the Universal Decay Law\cite{qi2009universal}, which is formulated within $R$-matrix theory to describe both cluster radioactivity and $\alpha$ decay. The second approach treats $\alpha$ decay as a microscopic quantum tunneling process. Gurvitz and Kalbermann noted that, because the $\alpha$-decay width is much smaller than the decay energy, the process can be well approximated by treating the $\alpha$ particle as a bound state in the internal nuclear potential and a scattering state in the external potential. Based on this physical picture, several semiclassical frameworks have been established, including the modified generalized liquid drop model (MGLDM)\cite{guo2015nuclear,zhang2006alpha}, the Coulomb and proximity potential model\cite{zanganah2020calculation,yahya2020alpha}, the two-potential approach (TPA)\cite{gurvitz1987decay},the Gamow-like model\cite{zdeb2013half}, and  the double-folding potential model \cite{moghaddari2020influence}.
\par{}
In recent years, significant efforts have been made to extend these empirical approaches. Notably, V. Yu. Denisov\cite{denisov2024empirical} introduced daughter-nucleus deformation parameters into an empirical framework for the first time, achieving substantial gains in describing $\alpha$-decay half-lives. G. Saxena et al\cite{saxena2024global}. formulated the NMTN expression, which incorporates the parent spin, centrifugal barrier effects, and contributions from unpaired nucleons. In a different development, Sobhani and Luo \cite{sobhani2025unified} proposed a unified empirical model for both $\alpha$ and $\beta$ decays; notably, the natural logarithm of the half-life scales linearly with $Q^{-1/4}$ (with $Q$ being the decay energy), in contrast to the traditional $Q^{-1/2}$ dependence in the Geiger–Nuttall systematics. G. Saxena et al.\cite{saxena2024theoretical} also developed the RRF formula within the generalized liquid drop model (GLDM) framework, featuring combined terms proportional to $Q^{-1/2}$ and $Q^{-1/4}$. These advances demonstrate that incorporating detailed nuclear structure inputs substantially improves the predictive power for $\alpha$-decay half-lives.
\par{}
In recent years, machine learning (ML) techniques have seen growing application in nuclear physics\cite{jyothish2025transfer,shree2025alpha,yuan2026machine,zhao2026explore,yang2026alpha,luo2025hybrid,li2022deep,ma2023simple,you2025nuclear}. In the study of $\alpha$-decay half-lives, ML approaches have primarily evolved along two directions. In the first approach, key observables governing $\alpha$ decay—such as the decay energy, orbital angular momentum, proton number, and daughter-nucleus deformation—are adopted as input features to directly train models for half-life predictions,\cite{ma2023simple} including those in the superheavy region. In the second approach, ML models are utilized to predict the $\alpha$-particle preformation factor ($P_\alpha$) by selecting relevant nuclear structure properties as inputs. The ML-predicted preformation factors are then integrated into half-life calculation frameworks,\cite{jalili2024decay} thereby markedly improving their predictive accuracy.
\par{}
In this work, we extend the empirical formula proposed by Sobhani and Luo for $\alpha$ decay by incorporating the angular momentum and daughter-nucleus deformation into its $Q^{-1/4}$ dependence on the decay energy. Using this modified empirical formula, we calculate the $\alpha$-decay half-lives of 400 nuclei categorized into even-even, even-odd, odd-even, and odd-odd groups. Furthermore, we establish a machine-learning model that takes as input features the primary physical quantities from the Sobhani formula (proton number, mass number, $Q^{-1/4}$, and its internal parameters) along with the newly included angular momentum and daughter deformation. The trained model is subsequently applied to evaluate the $\alpha$-decay half-lives of the same 400 nuclei. Finally, as a practical application, both the improved empirical formula and the trained machine-learning model are employed to predict the $\alpha$-decay half-lives of even-even superheavy nuclei with $Z=120$ and $122$, and the results are systematically compared with predictions from the original Sobhani and Poenaru formulas.
\par{} 
The paper is structured as follows. Section II introduces the modified empirical formulation for $\alpha$ decay and describes the implementation of the machine-learning models. Section III presents a detailed analysis of the calculations, highlighting the improvements introduced by the new empirical relation and evaluating the predictive capacity and constraints of the machine-learning framework. A brief summary and potential extensions of this work are given in Sec. IV.

\section{\label{sec:level2}FORMALISM OF \texorpdfstring{$\alpha$}{}-DECAY HALF-LIVES }
\begin{figure*}[ht]
    \centering
    \includegraphics[width=\textwidth]{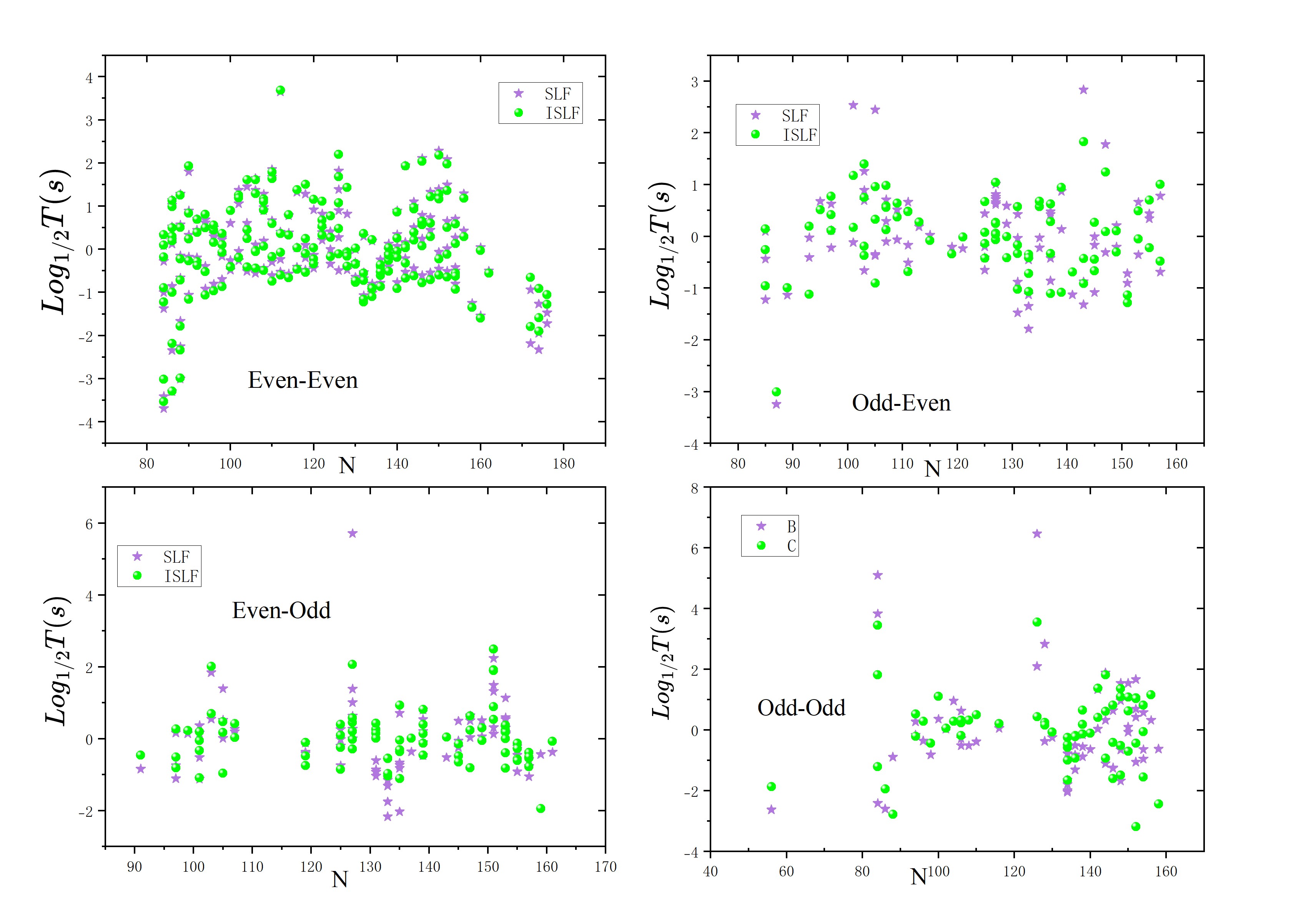}
    \caption{  Comparison of the $\alpha$-decay half-lives calculated using the SLF and ISLF empirical formulas. The vertical axis denotes the natural logarithm of the $\alpha$-decay half-life, $\ln(T_{1/2}/\text{s})$, and the horizontal axis represents the neutron number $N$ of the daughter/parent nuclei. Results from SLF are indicated by purple stars, while those from ISLF are depicted by green spheres.}
    \label{imag1}
\end{figure*}
\begin{figure*}[ht]
    \centering
    \includegraphics[width=\textwidth]{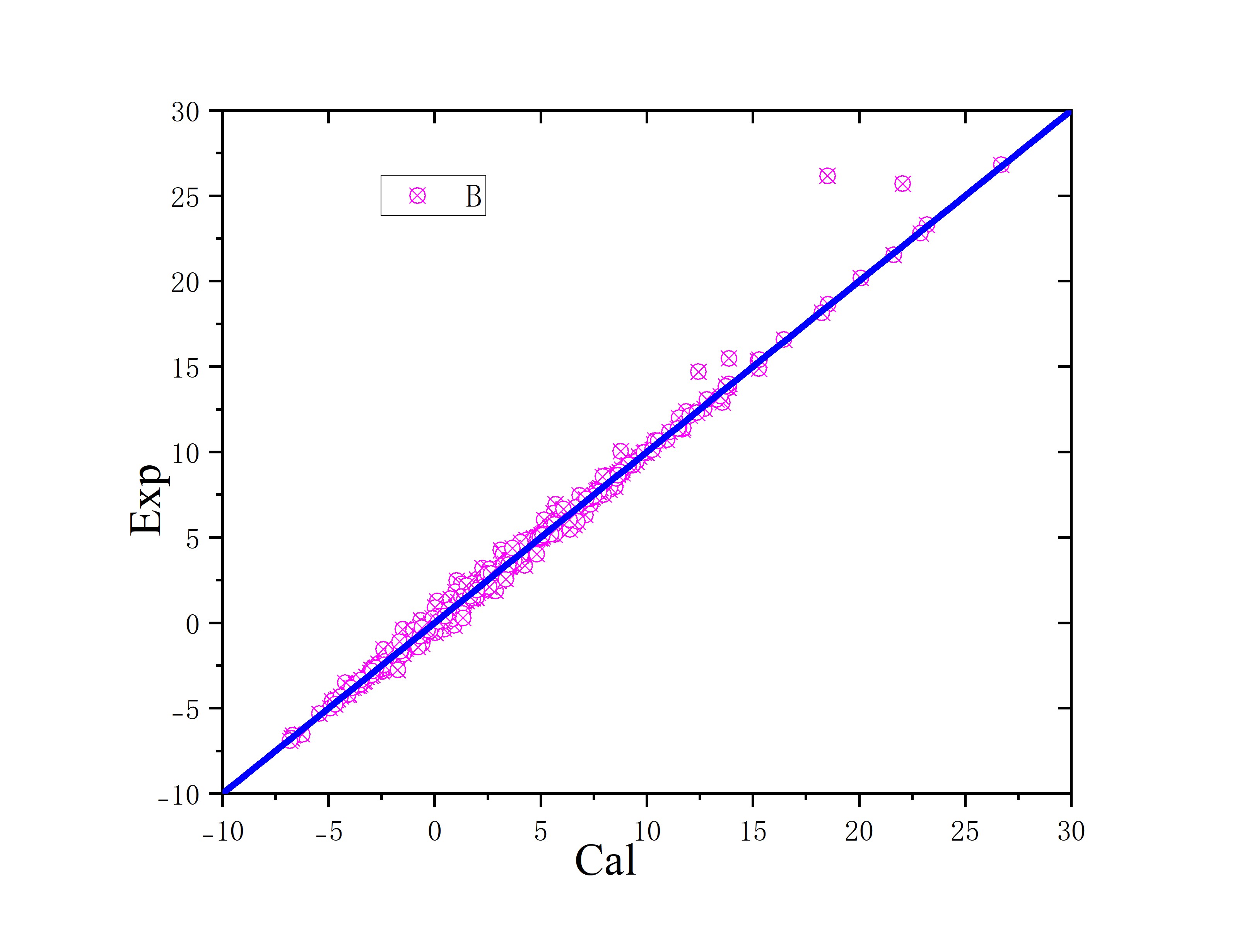}
    \caption{  Correlation plot of calculated versus experimental $\alpha$-decay half-lives, $\ln(T_{1/2}/\text{s})$. The horizontal and vertical axes show the SVRISLF predictions and experimental values, respectively. The diagonal line corresponds to $y = x$, representing perfect agreement between theory and experiment; individual nuclei are indicated by purple dots.}
    \label{imag2}
\end{figure*}
\begin{table*}[ht]
    \renewcommand{\arraystretch}{1}
    \setlength{\tabcolsep}{0.1cm}
    \centering
    \caption{The coefficient of the Sobhani and Luo Formula(SLF).}
    \begin{ruledtabular}
    \scalebox{1}{
    \begin{tabular}{cccccc}
        Set& a & b&c&d&e \\
        \colrule
        Even-even & 0.886731 &-0.216499 &161.126612&53.476122&-139.231272\\
        Even-odd & -0.625829 & 0.361162 &154.504828 &-65.580131&-103.636799\\
        Odd-even & 0.526851& -0.075369 & 159.282072&25.375995&-130.403663\\
        Odd-odd &0.418202 &-0.062085&137.374394&33.557035&-111.516774 \\
    \end{tabular}
    }
    \label{tab1}
    \end{ruledtabular}
\end{table*}
\begin{table*}[ht]
    \renewcommand{\arraystretch}{1}
    \setlength{\tabcolsep}{0.2cm}
    \centering
    \caption{The coefficient of the the Improved  Sobhani and Luo Formula(ISLF).}
    \begin{ruledtabular}
    \scalebox{1}{
    \begin{tabular}{cccccccc}
        Set& a & b&c&d&e&f&g \\
        \colrule
        Even-even & 0.897989 &-0.229347&156.946154& 57.269013& 1.000000&0.034056&-136.078788 \\
        Even-odd & -0.253083  & 0.251330 &177.996992  &-52.867385 &0.493424& -0.014029&-130.913952\\
        Odd-even & 0.454157& -0.018941&168.887515&5.189165&0.422338&-0.007241&-139.655842\\
        Odd-odd & 0.534682& -0.059948&162.244548&16.964812&0.644943&0.020992&-137.745482 \\
    \end{tabular}
    }
    \label{tab2}
    \end{ruledtabular}
\end{table*}
\begin{table*}[ht]
    \renewcommand{\arraystretch}{1}
    \setlength{\tabcolsep}{0.2cm}
    \centering
    \caption{The coefficient of the the SVRISLF.}
    \begin{ruledtabular}
    \scalebox{1}{
    \begin{tabular}{ccc}
        Set&MSE&$R^{2}$ \\
        \colrule
        All nucles &1.2384 &0.9611 \\
    \end{tabular}
    }
    \label{tab8}
    \end{ruledtabular}
\end{table*}
  \subsection{the improved formula }\label{A1}
\par{}
Recently, Sobhani and Luo propose \cite{sobhani2025unified} a unified empirical model for both $\alpha$ and $\beta$ decays. The new formalism is written as:
\begin{equation}\label{equal1}
\begin{aligned}
log_{10}T_{1/2}(s) =& aZ + bA \\&
                   +c Q^{-1/4}+d*I+e.                                                                                       
\end{aligned}
\end{equation}
where the half-life is given in seconds, the Z is the proton number, A is the mass number, Q is the $\alpha$-decay energy, and with $I=(N-Z)/(N+Z)$.
\par{}
In this work, we extend the Sobhani formula by incorporating the angular momentum and the deformation of the daughter nucleus, and propose an improved formula, which is given by

\begin{equation}\label{equal2}
\begin{aligned}
log_{10}T_{1/2}(s) =& aZ + bA \\
                   &+c Q^{-1/4}+d*I\\
                   &+e*\sqrt{l*(l+1)}+f(k\beta)^{1/2}\frac{Z}{Q^{1/2}}+g                                                                                     
\end{aligned}
\end{equation}
The values of the parameters in Eqs.(\ref{equal1})and (\ref{equal2})are given in Tables \ref{tab1} and \ref{tab2}, respectively.
\par{}
According to the WKB framework, the centrifugal barrier—determined by the orbital angular momentum of the emitted $\alpha$ particle—exerts a strong influence on the $\alpha$-decay half-life. In particular, this centrifugal effect contributes a term proportional to $\sqrt{l*(l+1)}$ to the logarithmic half-life. This provides a clear physical justification for introducing angular momentum into the refined empirical relationship.
\par{}
In 2024, Denisov \cite{denisov2024empirical} incorporated daughter-nucleus deformation into an empirical formula for $\alpha$ decay. Within the WKB approximation, the tunneling barrier is primarily determined by the Coulomb potential between the $\alpha$ particle and the daughter nucleus. Because many daughter nuclei exhibit quadrupole deformation rather than spherical symmetry, the minimum Coulomb barrier can be expressed as$$V_C^{\text{min}} = \frac{2(Z-2)e^2}{R_L + R_\alpha},$$where $R_L$ denotes the maximum radius of the deformed daughter nucleus and $R_\alpha$ is the radius of the $\alpha$ particle. Consequently, the modification of the tunneling barrier induced by nuclear deformation is proportional to $R_L - R_0$, where $R_0$ is the equivalent sharp radius of the spherical daughter nucleus. The angle-dependent radius of a quadrupole-deformed nucleus is parameterized as$$R(\theta) = R_0 \left[ 1 + \beta Y_{20}(\theta) \right].$$From this parameterization, the maximum radius $R_L$ depends explicitly on the sign of the quadrupole deformation parameter $\beta$. For prolate deformation ($\beta > 0$), the maximum radius occurs at $\theta = 0$, yielding$$R_L = R_0 \left( 1 + \sqrt{\frac{5}{4\pi}} \beta \right).$$Conversely, for oblate deformation ($\beta < 0$), the maximum radius occurs at $\theta = \pi/2$, given by$$R_L = R_0 \left( 1 - \frac{1}{2}\sqrt{\frac{5}{4\pi}} \beta \right).$$In Denisov's formulation, the deformation correction introduced into the empirical decay formula is expressed as
\begin{equation}\label{equal3}
\begin{aligned}
e(k\beta)^{1/2}\frac{Z}{Q^{1/2}}.                                                                                       
\end{aligned}
\end{equation}
\par{}
The Denisov \cite{denisov2024empirical} model provides a substantial improvement in predicting $\alpha$-decay half-lives for even-even isotopes, whereas the Sobhani systematics performs better across non-even-even systems. Consequently, by integrating the daughter-deformation parameterization from the Denisov formula and the centrifugal barrier effect into the Sobhani relation, we establish a refined empirical expression designed to achieve unified accuracy across all nuclear parity types.

\par{}
In the present study, the framework proposed by Sobhani and Luo is modified by explicitly accounting for both the centrifugal potential and daughter-nucleus quadrupole deformation following Denisov's approach. This enhanced model achieves an improved physical description of nuclear $\alpha$ decay and enhances the precision of half-life predictions across the nuclidic chart. The optimal model parameters were extracted using standard numerical curve-fitting routines.
 \subsection{machine learning}\label{B2}
\par{}
Machine learning fundamentally aims to map input-output relations from empirical data to predict outcomes for unseen instances. Here, we operate within the regression paradigm, where model parameters are adjusted to minimize discrepancies between predicted values and ground-truth labels in the training set. Specifically, we implement Support Vector Regression (SVR). The core concept of SVR relies on constructing a regression boundary governed by an $\epsilon$-insensitive zone of width $2\epsilon$, inside which deviations are not penalized. The formulation balances function flatness with the requirement of bounding most training instances within the $\epsilon$-tube. The primary advantage of SVR lies in the parameter $\epsilon$, which mitigates overfitting by accommodating minor deviations from training targets. Together with the regularization constant $C$, $\epsilon$ offers fine control over the bias-variance trade-off. This structural design not only reduces sensitivity to local statistical fluctuations but also improves extrapolation capability. Additionally, SVR exhibits strong performance in low-data regimes, making it highly suitable for nuclear physics applications where sample sizes are intrinsically constrained.
\par{}
To ensure robust performance, the hyperparameters of the SVR model were optimized using a grid search combined with 5-fold cross-validation (GridSearchCV). Because SVR is a mature regression framework, standard scikit-learn algorithmic implementations were adopted for hyperparameter tuning. This data-driven optimization protocol guarantees model stability and validates the overall feasibility of the proposed approach.
\begin{figure*}[b]
    \centering
    \includegraphics[width=\textwidth]{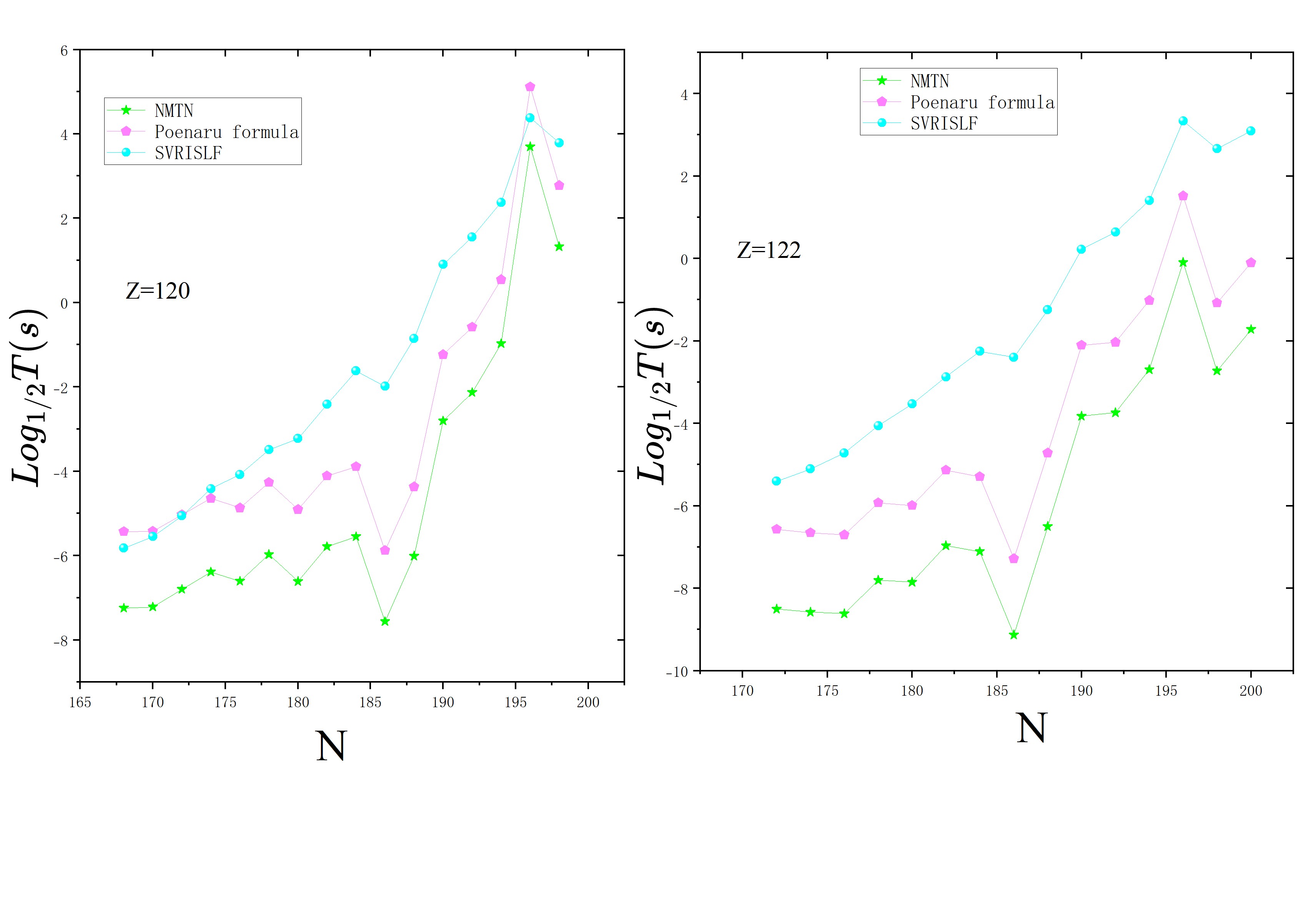}
    \caption{  Comparison of predicted $\alpha$-decay half-lives, $\ln(T_{1/2}/\text{s})$, for even-even superheavy nuclei with $Z = 120$ and $122$ versus neutron number $N$. Theoretical results from NMTN, SVRISLF, and the Poenaru formula are represented by green stars, purple pentagons, and blue circles, respectively.}
    \label{imag3}
\end{figure*}
\begin{figure*}[b]
    \centering
    \includegraphics[width=\textwidth]{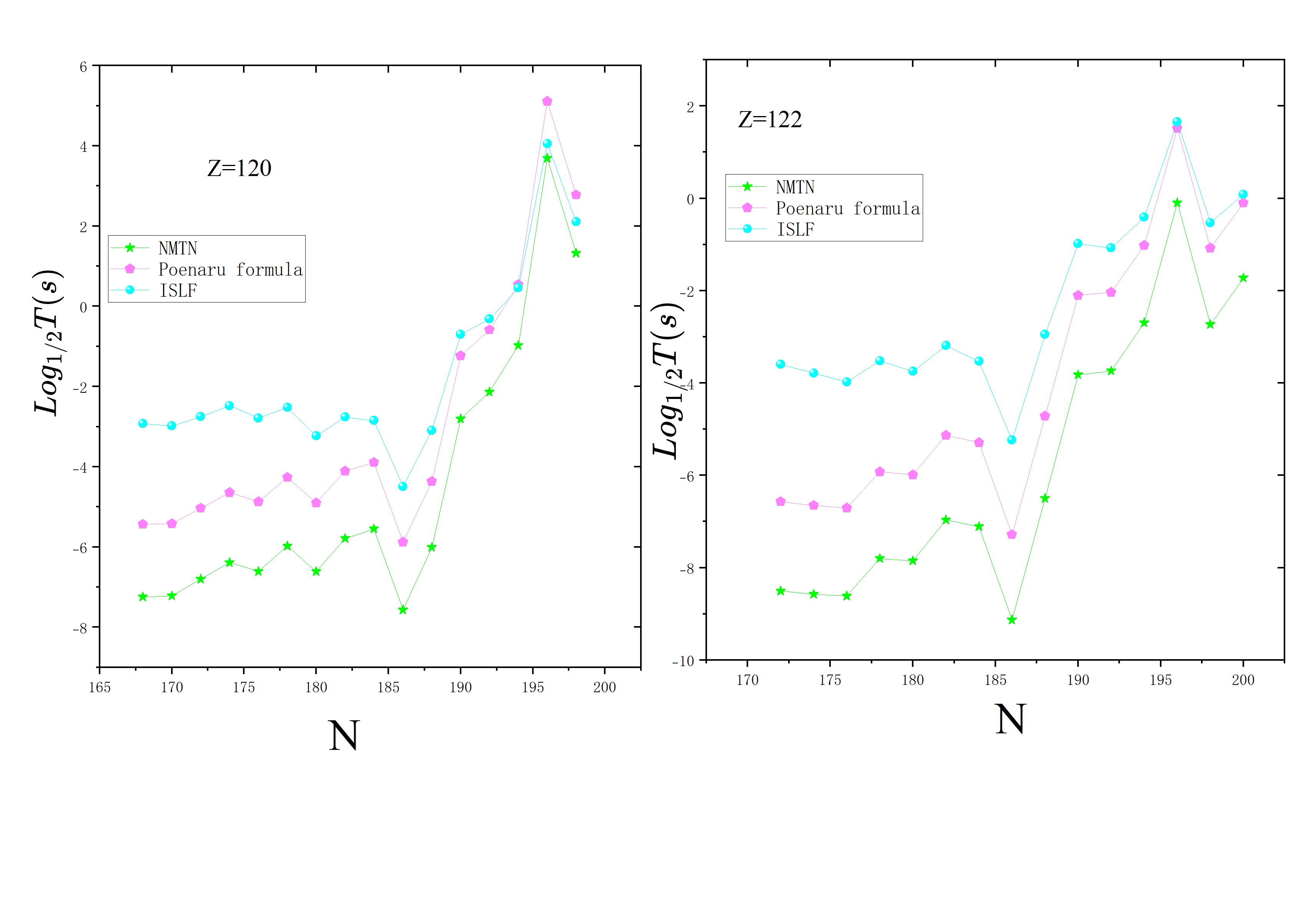}
    \caption{  Comparison of predicted $\alpha$-decay half-lives, $\ln(T_{1/2}/\text{s})$, for even-even superheavy nuclei with $Z = 120$ and $122$ versus neutron number $N$. Theoretical results from NMTN, ISLF, and the Poenaru formula are represented by green stars, purple pentagons, and blue circles, respectively.}
    \label{imag4}
\end{figure*}
\section{\label{sec:level3}RESULTS AND DISCUSSION}
In this work, Eq.(\ref{equal1})is denoted as SLF, and Eq(\ref{equal2}) as ISLF. The ISLF, excluding its constant term, is used as the input feature set for the support vector regression (SVR) model. The half‑lives of $\alpha$-decay for 400 nuclei are calculated using SLF, ISLF, and the SVR‑based variant (denoted SVRISLF). To evaluate the performance of these methods, the root‑mean‑square (RMS) deviations between the calculated and experimental half‑lives are computed, and the individual calculated values are also compared with the experimental data.The The RMS is written as
\begin{equation}\label{equa14}
\sigma = \left \{ \frac{1}{n} \sum_{i=1}^{n}\left [ log_{10}(T_{\frac{1}{2},i}^{calc.})-log_{10}(T_{\frac{1}{2},i}^{expt.}) \right ]^{2} \right \} ^{1/2}
\end{equation}
 where $log_{10}(T_{\frac{1}{2},i}^{calc.})$ and $log_{10}(T_{\frac{1}{2},i}^{expt.})$ are the calculated and experimental \texorpdfstring{$\alpha$}{}-decay half-lives the nucleus, and the $n$ is the number of nucleus.

\par{}
Table \ref{tab7} lists RMS errors of the SLF, ISLF, and SVRISLF calculations with respect to the corresponding experimental values.
\par{}
The difference between the calculated results and experimental data for each model can be expressed as:
\begin{equation}\label{equa15}
\Delta T = log_{10}(T_{\frac{1}{2},i}^{calc.})-log_{10}(T_{\frac{1}{2},i}^{expt.})
\end{equation}
\par{}
Similarly, the calculated $\alpha$‑decay half‑lives obtained using SLF, ISLF, and SVRISLF are given in Table \ref{tab8}, along with the differences relative to experiment. A detailed comparison between the ISLF and SLF results is presented in Fig. \ref{imag1}, providing a clearer visualization of the $\alpha$-decay half-lives. Furthermore, Fig. \ref{imag2} demonstrates that the SVRISLF model achieves the superior performance among the three approaches in describing nuclear $\alpha$ decay.
\par{}
In addition, we benchmarked our predictions against the approaches of Saxena et al. (the NMTN formula) and Poenaru et al. (the Poenaru formula) for unmeasured nuclei, thereby comprehensively evaluating the predictive capability of the present method.
\par{}
The NMTN formula \cite{saxena2024global} is given by:
\begin{equation}\label{equa6}
\begin{aligned}
log_{10}T_{1/2}(s) = & a\sqrt{\mu}(Z_{d}Q^{-1/2}-Z_{d}^{2/3})+b+c\sqrt{(l*(l+1))}  \\&
                   +d*\sqrt{(I*(I+1))}\\
\end{aligned}
\end{equation}
with, for $N>126$,  a = 0.7988,b = -18.3048, c = 0.3041, d = -3.0399, and h = 0.
\par{}
The Poenaru formula \cite{akrawy2017alpha} can written as:
\begin{equation}\label{equa6}
\begin{aligned}
log_{10}T_{1/2}(s) = & a+b*A^{1/6}*\sqrt{Z}+c*Z*Q^{-1/2}\\&
                   +d*I+e*I^{2}\\
\end{aligned}
\end{equation}
where the fitting parameters are as follows: a= -27.989, b= 0.940, c=1.532, d =-5.747, and e = 11.336.
\par{}
Lastly, the $\alpha$-decay half-lives of even-even superheavy isotopes with $Z = 120$ and $122$ were extrapolated using four models: NMTN, framework B, ISLF, and SVRISLF, with input decay energies supplied by the WS4 mass evaluation. The resulting half-lives are displayed in Figs. \ref{imag3}-\ref{imag4}. Integrated with the statistical analyses in Figs.\ref{imag1}-\ref{imag2} and Tables \ref{tab1}–\ref{tab2}, these evaluations confirm SVRISLF as the optimal predictive model. In addition, the $R^2$ metric in Table \ref{tab8} reflects the robust extrapolation reliability of SVRISLF with reduced overfitting risk. While ISLF exhibits high consistency with NMTN and framework B across the isotopic chain, SVRISLF converges toward these established models predominantly for $N \ge 186$. This behavior stems from SVRISLF inheriting the essential physical parameters of the ISLF expression. The fitted relation in Table \ref{tab7} reveals a substantial offset parameter between ISLF and SLF, which governs the predictive behavior of SVRISLF when extrapolated to unknown nuclides. Crucially, a distinct minimum in the calculated half-life curves is predicted by all four methods at $N_{d} = 184$, reinforcing the shell closure signature of $N_{d} = 184$ as a magic neutron number in the superheavy region.
\section{\label{sec:level4}SUMMARY AND CONCLUSION}
\par{}
Recently, Sobhani and Luo proposed an empirical relation for the statistical description of $\alpha$ and $\beta$ decays. In this work, we extend their formulation by explicitly incorporating the centrifugal potential and daughter-nucleus deformations, establishing the improved Sobhani–Luo formula (ISLF). Furthermore, the core physical quantities governing $\alpha$ decay in the ISLF are adopted as input features for a support vector regression (SVR) framework, designated as SVRISLF. Using the SLF, ISLF, and SVRISLF approaches, we systematically calculate the $\alpha$-decay half-lives of 400 nuclei across the nuclidic chart. Comparative analysis demonstrates that SVRISLF achieves the superior overall performance among the three evaluated methods.
\par{}
Additionally, both ISLF and SVRISLF were utilized to predict the $\alpha$-decay half-lives of 31 superheavy nuclei with $Z = 120$ and $122$, leveraging empirical systematics built from the $Z = 64\text{--}118$ region. While ISLF yields results in excellent agreement with the NMTN and B models, SVRISLF predictions converge toward these traditional frameworks predominantly for $N \ge 186$. This indicates that the extrapolation capability of the SVR approach for unobserved superheavy nuclides is slightly inferior to that of the conventional ISLF, NMTN, and B relations. Notably, the joint half-life systematics from all four methods reveal a pronounced dip near $N_{d} = 184$, reinforcing the shell closure assignment of $N_{d} = 184$ as the next magic neutron number.

\begin{table*}[b]
    \renewcommand{\arraystretch}{1}
    \setlength{\tabcolsep}{0.1cm}
    \centering
    \caption{The RMS deviation of the models SLF and ISLF.}
    \begin{ruledtabular}
    \scalebox{1}{
    \begin{tabular}{cccccc}
        Formula& Even-even n=181& Even-odd n=79&Odd-even n=80&Odd-odd n=60&All nucles \\
        \colrule
        SLF & 1.0511 & 1.0407&0.9260&1.64921\\
        ISLF & 1.0397 & 0.7546&0.7493&1.2470 \\
        SVRISLF & & & & &0.5650
    \end{tabular}
    }
    \label{tab7}
    \end{ruledtabular}
\end{table*}
\begin{table*}[b]
    \renewcommand{\arraystretch}{1}
    \setlength{\tabcolsep}{0.1cm}
    \centering
    \caption{$\Delta T$ different between  experimental and theoretical formulas.}
    \begin{ruledtabular}
    \scalebox{1}{
    \begin{tabular}{ccccccccc}
       \multirow{2}{*}{Formula} &\multicolumn{2}{c}{ Even-even} & \multicolumn{2}{c}{ Even-odd} &\multicolumn{2}{c}{ Odd-even} &\multicolumn{2}{c}{ Odd-odd}  \\
              \cmidrule(lr){2-3}\cmidrule(lr){4-5}\cmidrule(lr){6-7}\cmidrule(lr){8-9}
              & Minimum&Maximum & Minimum&Maximum & Minimum&Maximum & Minimum&Maximum \\
        \colrule
        SLF & -3.6917 & 3.6487&-2.1663&5.712 & -3.2480&2.8320&-6.4594&2.6255\\
        ISLF &  -3.6836 & 3.5324&-2.4923&1.9394& -1.8246 &3.0057&-3.5574&3.1842 \\
    \end{tabular}
    }
    \label{tab8}
    \end{ruledtabular}
\end{table*}
\begin{table*}[b]
    \renewcommand{\arraystretch}{1}
    \setlength{\tabcolsep}{0.1cm}
    \centering
    \caption{Calculated $\alpha$-decay half-lives for even-even superheavy nuclei with $Z = 120$ and $122$ across four different theoretical frameworks: NMTN, the Poenaru formula, SVRISLF, and ISLF. Input $\alpha$-decay energies ($Q_\alpha$) are adopted from the WS4 mass model \cite{wang2014surface}.}
    \begin{ruledtabular}
    \scalebox{1}{
    \begin{tabular}{lcccccc}
        Z& N& $Q_{\alpha}$&$log_{10}T_{NMTN}$&$log_{10}T_{Poenaru formula}$&$log_{10}T_{SVRISLF}$& $log_{10}T_{ISLF}$\\
        \colrule
        120 & 168 & 13.705 & -7.244330345 & -5.434324615 & -5.824842222 & -2.923542053 \\
120 & 170 & 13.676 & -7.216408998 & -5.423177685 & -5.552246409 & -2.981971032 \\
120 & 172 & 13.441 & -6.800799583 & -5.030901918 & -5.057445539 & -2.752399379 \\
120 & 174 & 13.215 & -6.390395094 & -4.643333547 & -4.416651797 & -2.478558063 \\
120 & 176 & 13.316 & -6.60995343  & -4.873893099 & -4.078727701 & -2.78480284  \\
120 & 178 & 12.981 & -5.975381198 & -4.265412741 & -3.491834742 & -2.523298967 \\
120 & 180 & 13.294 & -6.613658918 & -4.906195185 & -3.228001997 & -3.23402415  \\
120 & 182 & 12.866 & -5.788707316 & -4.110029863 & -2.414825785 & -2.763753136 \\
120 & 184 & 12.74  & -5.55299438  & -3.891965327 & -1.623352108 & -2.849528493 \\
120 & 186 & 13.765 & -7.565715185 & -5.880646464 & -1.988378835 & -4.492565458 \\
120 & 188 & 12.945 & -6.012460042 & -4.368451834 & -0.854835286 & -3.093425226 \\
120 & 190 & 11.478 & -2.806328898 & -1.233632795 &  0.901589562 & -0.698514252 \\
120 & 192 & 11.196 & -2.134431956 & -0.586063808 &  1.550662951 & -0.317552599 \\
120 & 194 & 10.739 & -0.975345023 &  0.539975356 &  2.370780996 &  0.457677644 \\
120 & 196 & 9.173  &  3.692514603 &  5.109868099 &  4.378831237 &  4.052791581 \\
120 & 198 & 9.912  &  1.323500122 &  2.773867869 &  3.787285909 &  2.103655093 \\
122 & 172 & 14.643 & -8.505623335 & -6.568489928 & -5.404920651 & -3.597441389 \\
122 & 174 & 14.67  & -8.576389874 & -6.653976445 & -5.10394456  & -3.783179588 \\
122 & 176 & 14.678 & -8.614311888 & -6.706788543 & -4.720057497 & -3.981455306 \\
122 & 178 & 14.197 & -7.803127942 & -5.925785414 & -4.055024888 & -3.517965526 \\
122 & 180 & 14.212 & -7.853058489 & -5.989566192 & -3.531987596 & -3.749074274 \\
122 & 182 & 13.714 & -6.966751312 & -5.134104485 & -2.871263926 & -3.190505374 \\
122 & 184 & 13.78  & -7.112289226 & -5.290917021 & -2.253760252 & -3.532231097 \\
122 & 186 & 14.918 & -9.129867006 & -7.284373818 & -2.395969494 & -5.238828569 \\
122 & 188 & 13.435 & -6.500771032 & -4.717617689 & -1.243813887 & -2.94818297  \\
122 & 190 & 12.141 & -3.822404898 & -2.102348035 &  0.221270808 & -0.984763232 \\
122 & 192 & 12.096 & -3.741178073 & -2.035206805 &  0.635527447 & -1.074860802 \\
122 & 194 & 11.638 & -2.693801665 & -1.01976459  &  1.406196935 & -0.410589944 \\
122 & 196 & 10.619 & -0.095424552 &  1.51765877  &  3.334950536 &  1.66142472  \\
122 & 198 & 11.637 & -2.730164731 & -1.078832137 &  2.661775458 & -0.528894423 \\
122 & 200 & 11.221 & -1.723189151 & -0.102131965 &  3.095780317 &  0.087868389 \\
    \end{tabular}
    }
    \label{tab9}
    \end{ruledtabular}
\end{table*}

\bibliography{study}

@article{qu2014comparative,
  title={Comparative studies of Coulomb barrier heights for nuclear models applied to sub-barrier fusion},
  author={Qu, WW and Zhang, GL and Zhang, HQ and Wolski, R},
  journal={Physical Review C},
  volume={90},
  number={6},
  pages={064603},
  year={2014},
  publisher={APS}
}

@article{sun2017systematic,
  title={Systematic study of $\alpha$ decay half-lives of doubly odd nuclei within the two-potential approach},
  author={Sun, Xiao-Dong and Deng, Jun-Gang and Xiang, Dong and Guo, Ping and Li, Xiao-Hua},
  journal={Physical Review C},
  volume={95},
  number={4},
  pages={044303},
  year={2017},
  publisher={APS}
}

@article{oganessian2010synthesis,
  title={Synthesis of a new element with atomic number Z= 117},
  author={Oganessian, Yu Ts and Abdullin, F Sh and Bailey, PD and Benker, DE and Bennett, ME and Dmitriev, SN and Ezold, Julie G and Hamilton, JH and Henderson, Roger A and Itkis, MG and others},
  journal={Physical review letters},
  volume={104},
  number={14},
  pages={142502},
  year={2010},
  publisher={APS}
}

@article{oganessian2007heaviest,
  title={Heaviest nuclei from 48Ca-induced reactions},
  author={Oganessian, Yuri},
  journal={Journal of Physics G: Nuclear and Particle Physics},
  volume={34},
  number={4},
  pages={R165},
  year={2007},
  publisher={IOP Publishing}
}

@article{hosseini2017theoretical,
  title={Theoretical approaches to alpha decay half-lives of super-heavy nuclei},
  author={Hosseini, SS and Hassanabadi, H},
  journal={Chinese Physics C},
  volume={41},
  number={6},
  pages={064101},
  year={2017},
  publisher={IOP Publishing}
}

@article{javadimanesh2013investigation,
  title={Investigation of deformed nuclei with a new potential combination},
  author={Javadimanesh, E and Hassanabadi, H and Rajabi, AA and Rahimov, H and Zarrinkamar, S},
  journal={Chinese Physics C},
  volume={37},
  number={11},
  pages={114102},
  year={2013},
  publisher={IOP Publishing}
}

@article{hosseini2019alpha,
  title={Alpha particle preformation factor of spherical nuclei for 67$\leq$ Z$\leq$91},
  author={Hosseini, SS and Hassanabadi, H and Akrawy, Dashty T},
  journal={Modern Physics Letters A},
  volume={34},
  number={05},
  pages={1950039},
  year={2019},
  publisher={World Scientific}
}

@article{hodgson2003cluster,
  title={Cluster emission, transfer and capture in nuclear reactions},
  author={Hodgson, Peter Edward and B{\v{e}}t{\'a}k, E},
  journal={Physics reports},
  volume={374},
  number={1},
  pages={1--89},
  year={2003},
  publisher={Elsevier}
}

@article{gamow1928quantentheorie,
  title={Zur quantentheorie des atomkernes},
  author={Gamow, George},
  journal={Zeitschrift f{\"u}r Physik},
  volume={51},
  number={3},
  pages={204--212},
  year={1928},
  publisher={Springer}
}

@article{geiger1911lvii,
  title={LVII. The ranges of the $\alpha$ particles from various radioactive substances and a relation between range and period of transformation},
  author={Geiger, Hans and Nuttall, JM},
  journal={The London, Edinburgh, and Dublin Philosophical Magazine and Journal of Science},
  volume={22},
  number={130},
  pages={613--621},
  year={1911},
  publisher={Taylor \& Francis}
}

@article{santhosh2018alpha,
  title={$\alpha$-decay chains of superheavy nuclei with Z= 125},
  author={Santhosh, KP and Nithya, C},
  journal={Physical Review C},
  volume={97},
  number={4},
  pages={044615},
  year={2018},
  publisher={APS}
}

@article{akrawy2019alpha,
  title={$\alpha$-decay systematics for superheavy nuclei},
  author={Akrawy, Dashty T and Ahmed, Ali H},
  journal={Physical Review C},
  volume={100},
  number={4},
  pages={044618},
  year={2019},
  publisher={APS}
}

@article{royer2000alpha,
  title={Alpha emission and spontaneous fission through quasi-molecular shapes},
  author={Royer, Guy},
  journal={Journal of Physics G: Nuclear and Particle Physics},
  volume={26},
  number={8},
  pages={1149},
  year={2000},
  publisher={IOP Publishing}
}

@article{qi2009universal,
  title={Universal decay law in charged-particle emission and exotic cluster radioactivity},
  author={Qi, Chong and Xu, FR and Liotta, Roberto J and Wyss, Ramon},
  journal={Physical review letters},
  volume={103},
  number={7},
  pages={072501},
  year={2009},
  publisher={APS}
}

@article{viola1966nuclear,
  title={Nuclear systematics of the heavy elements—II Lifetimes for alpha, beta and spontaneous fission decay},
  author={Viola Jr, VE and Seaborg, GT},
  journal={Journal of Inorganic and Nuclear Chemistry},
  volume={28},
  number={3},
  pages={741--761},
  year={1966},
  publisher={Elsevier}
}

@article{denisov2024empirical,
  title={Empirical relations for $\alpha$-decay half-lives: The effect of deformation of daughter nuclei},
  author={Denisov, V Yu},
  journal={Physical Review C},
  volume={110},
  number={1},
  pages={014604},
  year={2024},
  publisher={APS}
}

@article{zdeb2013half,
  title={Half-lives for $\alpha$ and cluster radioactivity within a Gamow-like model},
  author={Zdeb, A and Warda, M and Pomorski, K},
  journal={Physical Review C—Nuclear Physics},
  volume={87},
  number={2},
  pages={024308},
  year={2013},
  publisher={APS}
}

@article{guo2015nuclear,
  title={The nuclear deformation and the preformation factor in the $\alpha$-decay of heavy and superheavy nuclei},
  author={Guo, Shuqing and Bao, Xiaojun and Gao, Yuan and Li, Junqing and Zhang, Hongfei},
  journal={Nuclear Physics A},
  volume={934},
  pages={110--120},
  year={2015},
  publisher={Elsevier}
}

@article{zhang2006alpha,
  title={$\alpha$ decay half-lives of new superheavy nuclei within a generalized liquid drop model},
  author={Zhang, Hongfei and Zuo, Wei and Li, Junqing and Royer, Guy},
  journal={Physical Review C—Nuclear Physics},
  volume={74},
  number={1},
  pages={017304},
  year={2006},
  publisher={APS}
}

@article{zanganah2020calculation,
  title={Calculation of $\alpha$-decay and cluster half-lives for 197--226fr using temperature-dependent proximity potential model},
  author={Zanganah, V and Akrawy, Dashty T and Hassanabadi, H and Hosseini, SS and Thakur, Shagun},
  journal={Nuclear Physics A},
  volume={997},
  pages={121714},
  year={2020},
  publisher={Elsevier}
}

@article{yahya2020alpha,
  title={Alpha decay half-lives of 171-189Hg isotopes using Modified Gamow-like model and temperature dependent proximity potential},
  author={Yahya, WA},
  journal={Journal of the Nigerian Society of Physical Sciences},
  pages={250--256},
  year={2020}
}

@article{gurvitz1987decay,
  title={Decay width and the shift of a quasistationary state},
  author={Gurvitz, SA and Kalbermann, G},
  journal={Physical review letters},
  volume={59},
  number={3},
  pages={262},
  year={1987},
  publisher={APS}
}

@article{moghaddari2020influence,
  title={Influence of the Pauli exclusion principle on $\alpha$ decay},
  author={Moghaddari Amiri, M and Ghodsi, ON},
  journal={Physical Review C},
  volume={102},
  number={5},
  pages={054602},
  year={2020},
  publisher={APS}
}

@article{wang2014surface,
  title={Surface diffuseness correction in global mass formula},
  author={Wang, Ning and Liu, Min and Wu, Xizhen and Meng, Jie},
  journal={Physics Letters B},
  volume={734},
  pages={215--219},
  year={2014},
  publisher={Elsevier}
}

@article{sobhani2025unified,
  title={A unified formula for the half-life of the $\alpha$ and $\beta$ decay},
  author={Sobhani, Hadi and Luo, Yan-An},
  journal={Scientific Reports},
  volume={15},
  number={1},
  pages={41759},
  year={2025},
  publisher={Nature Publishing Group UK London}
}

@article{saxena2024global,
  title={A global study of $\alpha$-clusters decay in heavy and superheavy nuclei with half-life and preformation factor},
  author={Saxena, G and Sharma, PK and Saxena, Prafulla},
  journal={The European Physical Journal A},
  volume={60},
  number={3},
  pages={50},
  year={2024},
  publisher={Springer}
}

@article{saxena2024theoretical,
  title={Theoretical investigation of heavy cluster decay from Z= 118 and 120 isotopes: A search for an empirical formula in superheavy region},
  author={Saxena, G and Akrawy, Dashty T and Ahmed, Ali H and Aggarwal, Mamta},
  journal={Nuclear Physics A},
  volume={1046},
  pages={122867},
  year={2024},
  publisher={Elsevier}
}

@article{gurney1928wave,
  title={Wave mechanics and radioactive disintegration},
  author={Gurney, Ronald W and Condon, Edw U},
  journal={Nature},
  volume={122},
  number={3073},
  pages={439--439},
  year={1928},
  publisher={Nature Publishing Group UK London}
}

@article{jyothish2025transfer,
  title={Transfer-learning-driven machine-learning model for $\alpha$-decay half-life predictions},
  author={Jyothish, K and Manangode, Govardhan and Rhine Kumar, AK},
  journal={Physical Review C},
  volume={112},
  number={6},
  pages={064309},
  year={2025},
  publisher={APS}
}

@article{shree2025alpha,
  title={$\alpha$-decay half-life predictions for superheavy elements through machine learning techniques},
  author={Shree, S Madhumitha and Balasubramaniam, M},
  journal={The European Physical Journal A},
  volume={61},
  number={2},
  pages={32},
  year={2025},
  publisher={Springer}
}

@article{yuan2026machine,
  title={Machine learning-driven high-precision model for $\alpha$-decay energy and half-life prediction of superheavy nuclei},
  author={Yuan, Qingning and Qi, Panpan and Xiao, Xuanpeng and Wang, Xue and He, Juan and Long, Guimei and Duan, Zhengwei and Dai, Yangyan and Yan, Runchao and Yu, Gongming and others},
  journal={Physica Scripta},
  volume={101},
  number={17},
  pages={176004},
  year={2026},
  publisher={IOP Publishing}
}

@article{zhao2026explore,
  title={Explore the high precision and interpretability of deep learning in the calculation of $\alpha$ decay half life},
  author={Zhao, Tian Liang and Bao, Xiao Jun},
  journal={Physics Letters B},
  volume={879},
  pages={140609},
  year={2026},
  publisher={Elsevier}
}

@article{yang2026alpha,
  title={$\alpha$-decay half-lives of superheavy nuclei with support-vector regression},
  author={Yang, Haitao and Li, Xiaopan and Song, Xiefei and Ma, Dianxu and Yu, Gongming and Bao, Xiaojun},
  journal={Physical Review C},
  volume={113},
  number={1},
  pages={014307},
  year={2026},
  publisher={APS}
}

@article{luo2025hybrid,
  title={Hybrid neural network method of a multilayer perceptron and autoencoder for the $\alpha$-particle preformation factor in $\alpha$-decay theory},
  author={Luo, Jiaqi and Xu, Yang and Li, Xiaolong and Wang, Junxiang and Zhang, Yangjie and Deng, Jungang and Zhang, Fang and Ma, Nana},
  journal={Physical Review C},
  volume={111},
  number={3},
  pages={034330},
  year={2025},
  publisher={APS}
}

@article{li2022deep,
  title={Deep learning approach to nuclear masses and $\alpha$-decay half-lives},
  author={Li, Chen-Qi and Tong, Chao-Nan and Du, Hong-Jing and Pang, Long-Gang},
  journal={Physical Review C},
  volume={105},
  number={6},
  pages={064306},
  year={2022},
  publisher={APS}
}

@article{ma2023simple,
  title={Simple deep-learning approach for $\alpha$-decay half-life studies},
  author={Ma, Na-Na and Zhao, Tian-Liang and Wang, Wen-Xia and Zhang, Hong-Fei},
  journal={Physical Review C},
  volume={107},
  number={1},
  pages={014310},
  year={2023},
  publisher={APS}
}

@article{jalili2024decay,
  title={-decay half-life predictions with support vector machine},
  author={Jalili, Amir and Pan, Feng and Draayer, Jerry P and Chen, Ai-Xi and Ren, Zhongzhou},
  journal={Scientific Reports},
  volume={14},
  number={1},
  pages={30776},
  year={2024},
  publisher={Nature Publishing Group UK London}
}

@article{you2025nuclear,
  title={Nuclear deformation effects on $\alpha$-decay half-lives with empirical formula and machine learning: H. You et al.},
  author={You, Hong-Qiang and He, Xiao-Tao and Wu, Ren-Hang and Zhang, Shuang-Shuang and Li, Jing-Jing and He, Qing-Hua and Zhang, Hai-Qian},
  journal={Nuclear Science and Techniques},
  volume={36},
  number={10},
  pages={191},
  year={2025},
  publisher={Springer}
}

@article{akrawy2017alpha,
  title={Alpha decay calculations with a new formula},
  author={Akrawy, Dashty T and Poenaru, DN},
  journal={Journal of Physics G: Nuclear and Particle Physics},
  volume={44},
  number={10},
  pages={105105},
  year={2017},
  publisher={IOP Publishing}
}
\end{document}